\documentclass{article}

     \PassOptionsToPackage{numbers, compress}{natbib}
\usepackage[main,final,nonatbib]{neurips_2026}

\makeatletter
\renewcommand{\@trackname}{Conference on Physics and AI at Stanford University (PAI 2026).}
\makeatother

\usepackage[utf8]{inputenc} % allow utf-8 input
\usepackage[T1]{fontenc}    % use 8-bit T1 fonts
\usepackage{hyperref}       % hyperlinks
\usepackage{url}            % simple URL typesetting
\usepackage{booktabs}       % professional-quality tables
\usepackage{amsfonts}       % blackboard math symbols
\usepackage{nicefrac}       % compact symbols for 1/2, etc.
\usepackage{microtype}      % microtypography
\usepackage{xcolor}         % colors
\usepackage{cite}

\usepackage{comment}

\usepackage{amsmath}
\usepackage{graphicx}
\usepackage{adjustbox} % for max width tables
\usepackage{array}
\usepackage{tikz}
\usetikzlibrary{positioning, arrows.meta}
\usepackage{caption}
\usepackage{wrapfig}
\usepackage{tikz}
\newcommand{\gOneFourTikz}{%
\begin{tikzpicture}[
    baseline={([yshift=.3ex]current bounding box.south)},
    scale=0.2,
    every node/.style={circle,fill=black,inner sep=0.9pt},
    edge/.style={black,thin,line cap=round,line join=round}
]
    \coordinate (TL) at (0,1);
    \coordinate (TR) at (1.15,1);
    \coordinate (BL) at (0,0);
    \coordinate (BR) at (1.15,0);

    \draw[edge] (TL)--(TR);
    \draw[edge] (TL)--(BL);
    \draw[edge] (BL)--(BR);

    \node at (TL) {};
    \node at (TR) {};
    \node at (BL) {};
    \node at (BR) {};
\end{tikzpicture}%
}

\newcommand{\gThreeFourTikz}{%
\begin{tikzpicture}[
    baseline={([yshift=.3ex]current bounding box.south)},
    scale=0.2,
    every node/.style={circle,fill=black,inner sep=0.9pt},
    edge/.style={black,thin,line cap=round,line join=round}
]
    \coordinate (TL) at (0,1);
    \coordinate (TR) at (1.15,1);
    \coordinate (BL) at (0,0);
    \coordinate (BR) at (1.15,0);

    \draw[edge] (TL)--(TR)--(BR)--(BL)--cycle;

    \node at (TL) {};
    \node at (TR) {};
    \node at (BL) {};
    \node at (BR) {};
\end{tikzpicture}%
}

\newcommand{\gFourFourTikz}{%
\begin{tikzpicture}[
    baseline={([yshift=.3ex]current bounding box.south)},
    scale=0.2,
    every node/.style={circle,fill=black,inner sep=0.9pt},
    edge/.style={black,thin,line cap=round,line join=round}
]
    \coordinate (TL) at (0,1);
    \coordinate (TR) at (1.15,1);
    \coordinate (BL) at (0,0);
    \coordinate (BR) at (1.15,0);

    \draw[edge] (TL)--(TR);
    \draw[edge] (TL)--(BL);
    \draw[edge] (BL)--(BR);
    \draw[edge] (BL)--(TR);

    \node at (TL) {};
    \node at (TR) {};
    \node at (BL) {};
    \node at (BR) {};
\end{tikzpicture}%
}

\title{Graph Neural Networks for the Graphical Bootstrap }

\author{%
  Rigers Aliaj\\
  II. Institut f{\"u}r Theoretische Physik, Universit{\"a}t Hamburg\\
  Hamburg, Germany\\
  \texttt{rigers.aliaj@desy.de}
  \And
  Gabriele Dian\\
  Deutsches Elektronen-Synchrotron DESY\\
  Hamburg, Germany\\
  \texttt{dian.gabriele@gmail.com}
  \And
  Reza Doobary\\
  Independent Researcher\\
  United Kingdom\\
  \texttt{doobaryreza@gmail.com}
  \And
  Paul Heslop\\
  Department of Mathematical Sciences, Durham University\\
  Durham, United Kingdom\\
  \texttt{paul.heslop@durham.ac.uk}
}

\begin{document}

\maketitle

\begin{abstract}
 We study a graph classification problem involving over 20 million graphs, arising from high-order perturbative computations of correlators in planar $\mathcal{N}=4$ super-Yang--Mills, a model closely related to the theory of the strong nuclear force. We benchmark graph neural networks, including graph transformers, 
 achieving robust generalization to larger graphs with up to $99.996\%$ ROC AUC. Then, we analyze how the models can be used to gain a computational speedup compared to the traditional graphical bootstrap algorithm, through shrinking the redundant data by up to $85.5\%$ at the level of denominator graphs. Finally, we study the embeddings of the models to investigate their interpretability.
\end{abstract}

\section{Introduction }
With the upcoming high-luminosity upgrade of the LHC, particle-physics data are expected to reach unprecedented precision. To match this and further test our understanding of the underlying physics, comparably precise higher-order perturbative calculations are required. The object of interest is typically the scattering amplitude, whose perturbative expansion translates into diagrams with increasing number of vertices, while the number of such diagrams grows factorially with each perturbative order. This increase in complexity requires identifying new structures and simplifications. An ideal setting to develop and test such new ideas is $\mathcal{N}=4$ super Yang-Mills theory (SYM)~\cite{Henn:2020omi}.

In particular, we focus on 4-point correlator integrands (a simple limit of which yields amplitude integrands) in  planar $\mathcal{N}=4$ SYM. These have been computed to higher and higher orders in perturbation theory over many years~\cite{Eden:2000mv,Eden:2011we,Eden:2012tu,Bourjaily:2015bpz,Bourjaily:2016evz,He:2024cej} culminating in the recent remarkable 12-th order result~\cite{Bourjaily:2025iad}.  The four point integrand at $\ell$-th perturbative order, is a permutation invariant rational function of $n=4+{\ell}$ spacetime points $x_{1},\ldots,x_{n}$, allowing for an ansatz expansion in purely combinatorial objects, 
\begin{equation}
\label{eq:ansatz}
    f^{(n)}(x_1,\ldots,x_{n}) = \sum_{a,b} c^{(n)}_{a,b} \textcolor{blue}{f^{(n)}_{a,b}}(x_1,\ldots,x_{n}) \,.
\end{equation}
 Here the  $c^{(n)}_{a,b}$ are numerical coefficients and  $\textcolor{blue}{f^{(n)}_{a,b}}$ are permutation invariant functions of $x_i$. Strong physical requirements allow us to associate these functions with  graphs, called $f$-graphs~\cite{Eden:2011we,Eden:2012tu}, which have $n$ nodes and two types of edges, numerator edges, $N(f_{a,b}^{(n)})$, and denominator edges, $D(f_{a,b}^{(n)})$. The denominator edges alone must form a planar graph, and the net (denominator minus numerator) valency at each vertex must be four.
 The precise relation is
\begin{equation}\label{fgraphs}
\textcolor{blue}{f^{(n)}_{a,b}}(x_1,\ldots,x_{n}) = \sum_{\sigma \in S_{n}} \frac 1{|\text{Aut}(f^{(n)}_{a,b})|} \frac{\prod_{(ij)\in N(f^{(n)}_{a,b})} x_{ij}^2}{\prod_{(ij) \in D(f^{(n)}_{a,b})} x_{ij}^2}\,. 
\end{equation}
Here  $x_i$ only appears through distance pairs $x_{ij}^2:=(x_{i}-x_{j})^2$,   $|\text{Aut}(f_{a,b}^{(n)})|$ is the graph symmetry factor and  $S_{n}$ represents the $n$-point permutation group. 
The full set of relevant f-graphs is generated by first drawing all physically allowed denominator graphs and then dressing them with numerator edges in all possible ways. The label $a$ runs over the various denominator graphs (black lines) while $b$, whose relevance only begins at $n=10$, the different dressings with numerator edges (blue, dashed lines). For example at 8 points in figure~\ref{fig:f-graphs} there are only 3 possible $f$-graphs, the first two admit a single numerator decoration while the third one none. Here all coefficients have values $\pm1$, whereas at higher $n$, zero coefficients, higher integers and fractional coefficients appear. The statistics of $f$-graphs and denominator graphs ($d$-graphs) are shown in table~\ref{tab:graphs-numbers}.
\begin{table}[h] 
\centering
\begin{tabular}{c|cccccccc}
\toprule
\# nodes & $\leq 9$ & 10 & 11 & 12 & 13 & 14 & 15 & 16 \\
\midrule
\# $f$-graphs & 13 & 36 & 220 & 2,707 & 42,979 & 898,353 & 22,024,902 & 619,981,403 \\
\# $d$-graphs & 13 & 31 & 164 & 1,430 & 13,936 & 152,009 & 1,697,302 & 19,212,867 \\
\% Vanishing $c_{a,b}$ & 0 & 28 & 42 & 61 & 76 & 85 & 91 & 94 \\
\% Vanishing $d_{a}$ & 0 & 27 & 35 & 49 & 59 & 66 & 71 & 75 \\
\bottomrule
\end{tabular}
\caption{Number of $f$-, $d$-graphs and percentage of vanishing coefficients per node number.}
\label{tab:graphs-numbers}
\end{table}
\begin{wrapfigure}[16]{r}{0.45\textwidth}
\centering
%\documentclass[tikz,border=5pt]{standalone}
%\usepackage{amsmath}
%\usetikzlibrary{calc,decorations.pathreplacing}
%\begin{document}

\newcommand{\fgraphone}{\vcenter{\hbox{%
\begin{tikzpicture}[
    baseline={(current bounding box.center)},
    scale=0.4,
    every node/.style={circle,fill=black,inner sep=1pt},
    baseedge/.style={black,thin},
    dashededge/.style={blue,thin,dashed}
]
    \coordinate (A) at (0,3.6);
    \coordinate (B) at (0,0);
    \coordinate (C) at (1.55,2.25);
    \coordinate (D) at (0.95,1.25);
    \coordinate (E) at (0.32,1.70);
    \coordinate (F) at (0.88,2.45);
    \coordinate (G) at (1.58,1.32);
    \coordinate (H) at (3.10,1.75);

    \draw[baseedge] (A)--(B);
    \draw[baseedge] (A)--(H);
    \draw[baseedge] (B)--(H);
    \draw[baseedge] (A)--(E);
    \draw[baseedge] (A)--(F);
    \draw[baseedge] (A)--(C);
    \draw[baseedge] (B)--(E);
    \draw[baseedge] (B)--(D);
    \draw[baseedge] (B)--(G);
    \draw[baseedge] (E)--(F);
    \draw[baseedge] (E)--(D);
    \draw[baseedge] (F)--(D);
    \draw[baseedge] (F)--(C);
    \draw[baseedge] (D)--(C);
    \draw[baseedge] (D)--(G);
    \draw[baseedge] (C)--(G);
    \draw[baseedge] (C)--(H);
    \draw[baseedge] (G)--(H);

    \draw[dashededge] (A) to[bend left=1] (D);
    \draw[dashededge] (B) to[bend right=16] (C);

    \node at (A) {};
    \node at (B) {};
    \node at (C) {};
    \node at (D) {};
    \node at (E) {};
    \node at (F) {};
    \node at (G) {};
    \node at (H) {};
\end{tikzpicture}%
}}}

\newcommand{\fgraphtwo}{\vcenter{\hbox{%
\begin{tikzpicture}[
    baseline={(current bounding box.center)},
    scale=0.4,
    every node/.style={circle,fill=black,inner sep=1pt},
    baseedge/.style={black,thin},
    dashededge/.style={blue,thin,dashed}
]
    \coordinate (A) at (0,3.6);
    \coordinate (B) at (0,0);
    \coordinate (U) at (0.35,2.26);
    \coordinate (C) at (1.33,2.35);
    \coordinate (L) at (0.55,1.46);
    \coordinate (M) at (0.96,1.22);
    \coordinate (R) at (1.77,1.43);
    \coordinate (H) at (3.10,1.77);

    \draw[baseedge] (A)--(B);
    \draw[baseedge] (A)--(H);
    \draw[baseedge] (B)--(H);
    \draw[baseedge] (A)--(U);
    \draw[baseedge] (A)--(C);
    \draw[baseedge] (B)--(U);
    \draw[baseedge] (B)--(L);
    \draw[baseedge] (B)--(M);
    \draw[baseedge] (U)--(L);
    \draw[baseedge] (U)--(C);
    \draw[baseedge] (L)--(M);
    \draw[baseedge] (L)--(C);
    \draw[baseedge] (M)--(C);
    \draw[baseedge] (M)--(R);
    \draw[baseedge] (M)--(H);
    \draw[baseedge] (C)--(R);
    \draw[baseedge] (C)--(H);
    \draw[baseedge] (R)--(H);

    \draw[dashededge] (C) to[bend left=30] (B);
    \draw[dashededge] (C) to[bend right=30] (B);

    \node at (A) {};
    \node at (B) {};
    \node at (U) {};
    \node at (C) {};
    \node at (L) {};
    \node at (M) {};
    \node at (R) {};
    \node at (H) {};
\end{tikzpicture}%
}}}

\newcommand{\fgraphthree}{\vcenter{\hbox{%
\begin{tikzpicture}[
    baseline={(current bounding box.center)},
    scale=0.4,
    every node/.style={circle,fill=black,inner sep=1pt},
    baseedge/.style={black,thin},
    dashededge/.style={blue,thin,dashed}
]
    \coordinate (L)  at (0,1.7);
    \coordinate (T)  at (1.2,3.3);
    \coordinate (R)  at (2.4,1.7);
    \coordinate (Bo) at (1.2,0.1);
    \coordinate (UL) at (0.82,2.05);
    \coordinate (UR) at (1.58,2.05);
    \coordinate (DL) at (0.82,1.25);
    \coordinate (DR) at (1.58,1.25);

    \draw[baseedge] (L)--(T)--(R)--(Bo)--(L);
    \draw[baseedge] (L)--(UL);
    \draw[baseedge] (L)--(DL);
    \draw[baseedge] (T)--(UL);
    \draw[baseedge] (T)--(UR);
    \draw[baseedge] (R)--(UR);
    \draw[baseedge] (R)--(DR);
    \draw[baseedge] (Bo)--(DL);
    \draw[baseedge] (Bo)--(DR);
    \draw[baseedge] (UL)--(UR);
    \draw[baseedge] (UL)--(DL);
    \draw[baseedge] (UR)--(DR);
    \draw[baseedge] (DL)--(DR);

    \node at (L) {};
    \node at (T) {};
    \node at (R) {};
    \node at (Bo) {};
    \node at (UL) {};
    \node at (UR) {};
    \node at (DL) {};
    \node at (DR) {};
\end{tikzpicture}%
}}}

\begin{center}
\resizebox{0.98\linewidth}{!}{%
\begin{tabular}{c c c c c c}
$\displaystyle f^{(8)}=$ & $\fgraphone$ & $+$ & $\fgraphtwo$ & $-$ & $\fgraphthree$ \\
& {\small $f_{1,1}^{(8)}$} & & {\small $f_{2,1}^{(8)}$} & & {\small $f_{3,1}^{(8)}$}
\end{tabular}%
}
\end{center}

%\end{document}
\caption{Expansion of $f^{(8)}$  in terms of f-graphs.}
\label{fig:f-graphs}
\begin{tikzpicture}[
   scale=0.5,
    every node/.style={circle,fill=black,inner sep=1.0pt},
    baseedge/.style={black,thin},
    edge/.style={thin}
]

\begin{scope}
    \coordinate (L)  at (0,1.7);
    \coordinate (T)  at (1.2,3.3);
    \coordinate (R)  at (2.4,1.7);
    \coordinate (Bo) at (1.2,0.1);
    \coordinate (UL) at (0.82,2.05);
    \coordinate (UR) at (1.58,2.05);
    \coordinate (DL) at (0.82,1.25);
    \coordinate (DR) at (1.58,1.25);

    \draw[baseedge] (L)--(T)--(R)--(Bo)--(L);
    \draw[baseedge] (L)--(UL);
    \draw[baseedge] (L)--(DL);
    \draw[baseedge] (T)--(UL);
    \draw[baseedge] (T)--(UR);
    \draw[baseedge] (R)--(UR);
    \draw[baseedge] (R)--(DR);
    \draw[baseedge] (Bo)--(DL);
    \draw[baseedge] (Bo)--(DR);
    \draw[red,edge] (UL)--(UR);
    \draw[red,edge] (UL)--(DL);
    \draw[red,edge] (UR)--(DR);
    \draw[red,edge] (DL)--(DR);

    \node at (L) {};
    \node at (T) {};
    \node at (R) {};
    \node at (Bo) {};
    \node at (UL) {};
    \node at (UR) {};
    \node at (DL) {};
    \node at (DR) {};
\end{scope}

\node[draw=none,fill=none,shape=rectangle,font=\large] at (4.4,1.75) {$\rightarrow$};

\begin{scope}[shift={(6.4,0)}]
    \coordinate (Lr)  at (0,1.7);
    \coordinate (Tr)  at (1.2,3.3);
    \coordinate (Rr)  at (2.4,1.7);
    \coordinate (Bor) at (1.2,0.1);
    \coordinate (ULr) at (0.82,2.05);
    \coordinate (URr) at (1.58,2.05);
    \coordinate (DLr) at (0.82,1.25);
    \coordinate (DRr) at (1.58,1.25);
    \coordinate (O) at (1.2,1.65);

    \draw[baseedge] (Lr)--(Tr)--(Rr)--(Bor)--(Lr);
    \draw[baseedge] (Lr)--(ULr);
    \draw[baseedge] (Lr)--(DLr);
    \draw[baseedge] (Tr)--(ULr);
    \draw[baseedge] (Tr)--(URr);
    \draw[baseedge] (Rr)--(URr);
    \draw[baseedge] (Rr)--(DRr);
    \draw[baseedge] (Bor)--(DLr);
    \draw[baseedge] (Bor)--(DRr);
    \draw[baseedge] (ULr)--(O);
    \draw[baseedge] (URr)--(O);
    \draw[baseedge] (DRr)--(O);
    \draw[baseedge] (DLr)--(O);
    \draw[red,edge] (ULr)--(URr);
    \draw[red,edge] (ULr)--(DLr);
    \draw[red,edge] (URr)--(DRr);
    \draw[red,edge] (DLr)--(DRr);
    \draw[blue,dash pattern=on 4pt off 2pt] (ULr) to[bend left=40] (DRr);
    \draw[blue,dash pattern=on 4pt off 2pt] (URr) to[bend right=40] (DLr);

    \node at (Lr) {};
    \node at (Tr) {};
    \node at (Rr) {};
    \node at (Bor) {};
    \node at (ULr) {};
    \node at (URr) {};
    \node at (DLr) {};
    \node at (DRr) {};
    \node at (O) {};
\end{scope}

\end{tikzpicture}
\caption{Two graphs related by the rung rule, which implies that their coefficients are equal.}
\label{fig:rung_rule}
\end{wrapfigure}
The computation of each perturbative term progresses by constructing an ansatz as in \eqref{eq:ansatz} using all allowed f-graphs. Physical constraints translate into graphical rules~\cite{Bourjaily:2016evz} (the graphical bootstrap) such as the ``rung rule'' (Figure~\ref{fig:rung_rule}), which replaces a squared face with a pyramid dressed with numerator edges on the diagonals, or the more powerful ``cusp rule'' \cite{He:2024cej,Bourjaily:2025iad}, providing linear equations among the various $c_{a,b}^{(n)}$. The rung rule is very simple but leaves many coefficients undetermined (14\% at $n=16$), whereas the cusp rule is more complete and may give a unique solution for all coefficients~\cite{He:2024cej}. 
However, as shown in Table \ref{tab:graphs-numbers}, the number of $f$-graphs grows factorially; at $n=16$, the cusp rule yields a system of approximately $10^{10}$ sparse equations that required 3 days on an HPC cluster to be solved \cite{Bourjaily:2025iad}. At $n > 16$, the construction and solution of such systems become computationally prohibitive.

One would like to use machine learning to predict the graph coefficients directly. However, as with all exact analytic results, ultimately 100\% accuracy is sought. Instead therefore, we propose a more modest goal which will have immediate practical use without needing such high  accuracy.
Although the number of graphs grows to a monstrous scale, exceeding 600 million at $n = 16$, the ansatz becomes increasingly redundant, with $94\%$ of $f$-graphs not contributing (table~\ref{tab:graphs-numbers}). 
Rather than predicting the coefficients themselves, we use machine learning to identify graphs that can be discarded $(c_{a,b}^{(n)}=0)$. 
We simplify further, by considering families  of graphs with the same underlying denominator together ($d$-graphs) and so predict
the binary coefficient:
\begin{equation} 
\label{eq:dengraph}
    d^{(n)}_a = \begin{cases} 0 & \text{if } c^{(n)}_{a,b}=0 \ \forall b \\ 1 & \text{otherwise} \end{cases}
\end{equation}
which simply states that if all numerator dressings of a denominator graph vanish,  it has label 0, otherwise 1.
This reduces the problem to a binary classification task on simple unlabelled planar graphs. 
 Since the system is solvable only when all graphs with coefficients equal to 1 are present, we require $100\%$ recall on positives. The reduction of the redundant fraction of the ansatz is then quantified by the True Negative Rate (TNR), which measures the fraction of correctly identified negative-class graphs.
\paragraph{Related work} 
The use of machine learning to obtain exact or symbolic results in theoretical high-energy physics has recently attracted growing attention, see for example~\cite{cai2024transforming}. This work employs Transformers to predict exact coefficient data in  planar $\mathcal{N}=4$ SYM correlators. Related directions include symbolic amplitude computation using machine learning~\cite{alnuqaydan2023symba} and the simplification of polylogarithmic expressions~\cite{dersy2022simplifying}. 

\section{Models, features and methodology}
We benchmark three architectures: Graph Isomorphism Network (GIN)~\cite{xu2019powerful}, Graph Attention Network (GAT)~\cite{velivckovic2018graph}, and Graphormer (GT)~\cite{Ying2021DoTR}. This enables a direct comparison between message passing, attention-based message passing, and transformer-style global attention within a unified setup.

All models share the same node inputs that combine three feature families: centrality scores (degree, clustering, betweenness, closeness, PageRank)~\cite{duong2019nodefeaturesgraphneural}, top-3 Laplacian eigenvectors as positional encodings~\cite{dwivedi2022graphneuralnetworkslearnable}, and graphlet orbit counts up to $k=4$~\cite{cleveland2023graphlet}. Since the coefficients are constrained by physical properties of the correlator that depend on graph substructures, we select features sensitive to local structural patterns.

GIN and GAT share a backbone of $L$ layers with batch normalization, Jumping Knowledge~\cite{xu2018representation}, and an MLP head; GIN uses sum pooling, GAT attention pooling with residuals. The GT uses a pre-LayerNorm Transformer~\cite{vaswani2017attention} with global self-attention and adjacency/shortest-path biases. 

Each experiment involves training on graphs with up to $n$ nodes and predicting coefficients for graphs with $n+1$ nodes, we refer to them as $n\!\to\! n+1$. Graphs with $n \leq 9$ have distinct statistical properties (no vanishing coefficients) so we exclude these. This binary classification task faces an increasing  distribution shift, as class imbalance grows with $n$
 (table~\ref{tab:graphs-numbers}).
 To ensure cross-size generalization, hyperparameters were selected by training on $10\le N\le n-1$ node graphs and validating on $n$-node graphs (i.e., a graph-size based split rather than a random partition).
 Then we ran a 300-point random grid search (learning rate $5\cdot10^{-5}-5\cdot10^{-3}$, weight decay $5\cdot10^{-7}-10^{-3}$, dropout 0.2--0.5,  2--5 layers, 4--128 hidden channels, batch size 4--256; GT heads 8--32) for 10 epochs. 
 Finally, models with the hyperparameters of the six best runs (top-3 by validation loss and top-3 by validation ROC AUC) are trained for 100 epochs on the full $10,\ldots,n$ data and we adopt the average of their raw outputs as per model predictions.  Optimization uses Adam, binary cross-entropy-with-logits, and OneCycle scheduling. Training runs on 1--4 GPUs (A100/V100 class), with the heaviest jobs taking 8--12 hours.
\section{Experiments, results and interpretability} 
\paragraph{Ranking and class separability}

Models are trained up to $12\leq n \leq 15$ node graphs and always tested on $n+1$ ones, matching the physics setting where $d_{a}^{(n+1)}$ are inferred from $d_{a}^{(n)}$. We are eventually interested in the ansatz reduction task (requiring 100\% recall), but it is initially interesting to consider the ranking of predicted probabilities among the two classes and analyse the choice of a threshold in the next paragraph.

To this end, Table~\ref{tab:roc_auc_binary} reports ROC AUC of the three architectures on different training setups. We observe a strong performance across all experiments , which improves, despite stronger class imbalance, at higher $n$. GT is consistently the best, reaching an impressive $99.996\%$, while GIN becomes competitive only when training includes at least $n=13$ data. The results strongly indicate that, indeed, the graphs constitute a learnable dataset, since despite the distribution shift between training and testing there is a strong threshold independent class separation.
\begin{table}[h]
\centering
\begin{tabular}{c c c c c}
\hline
 & 12 $\!\to\!$ 13 & 13 $\!\to\!$ 14 & 14 $\!\to\!$ 15 & 15 $\!\to\!$ 16 \\
\hline
 GIN & \textcolor{gray}{88.608} & \textcolor{gray}{95.647} & 98.534 & 98.600 \\
 \hline
 GAT & 89.597 & 95.679 & \textcolor{gray}{97.871} & \textcolor{gray}{98.137} \\
\hline
 GT & \textbf{97.921} & \textbf{99.459} & \textbf{99.959} & \textbf{99.996} \\
 \hline
\end{tabular}
\caption{ROC AUC (\%) performance for the different experiments and models.}
\label{tab:roc_auc_binary}
\end{table}

To interpret these results from a physics perspective, we ask how well the models capture the simplest among the graphical rules, the rung rule, which under the denominator-graph labelling in~\eqref{eq:dengraph}, only relates  graphs of positive class ($d_a=1$). To quantify the effect we compute the probability of superiority~\cite{mcgraw1992common} between rung-rule-generated data and its complement in the positive class across all experiments. GIN and GAT show weak-to-moderate separation ($50$--$55\%$ and $48$--$69\%$, respectively), whereas GT shows a stronger effect, increasing from $61\%$ to $86\%$ as training size grows. 
This indicates that GT captures rung-rule relevant substructures more effectively which is justified by the access that the Transformer architecture provides to complicated graph substructures.
\paragraph{Ansatz reduction}
Next, the ansatz reduction task is addressed with special focus on its operational constraint (100\% Recall) which is threshold-sensitive. We ask: is there a threshold satisfying the constraint that can be estimated robustly while also removing a substantial fraction of the ansatz?
We define $\mathbf{t_{\mathrm{train}}}$ ($\mathbf{t_{\mathrm{test}}}$) as the minimum predicted probability over true positives in the training (test) dataset for each model. Thus, $\mathbf{t_{\mathrm{test}}}$ is the optimal threshold: the smallest value that guarantees zero false negatives (FN). We observe that no predictable mapping from $\mathbf{t_{\mathrm{train}}}$ to $\mathbf{t_{\mathrm{test}}}$ exists across architectures and experiments. GAT is the most stable: $\mathbf{t_{\mathrm{train}}}/10$ guarantees zero false negatives  across all experiments while still hitting a high TNR, reaching $43.5, 50.4$ and $50\%$ for $n+1=14,15,16$ respectively. For GIN and GT, in contrast, the optimal threshold can lie several orders of magnitude below $\mathbf{t_{\mathrm{train}}}$ (Table~\ref{tab:ansatz_reduction}).
\begin{table}[h]
\centering
\begin{tabular}{@{}c c c c c c c@{}}
\hline
 & $\mathbf{t_{\mathrm{train}}}$ & $\mathbf{t_{\mathrm{test}}}$ & FN @ $\mathbf{t_{\mathrm{train}}}$ & FN @$\mathbf{t_{\mathrm{train}}}/10$ & TNR @$\mathbf{t_{\mathrm{train}}}/10$ & TNR @$\mathbf{t_{\mathrm{test}}}$ \\
\hline
GIN & 0.026 & $1.14\cdot10^{-5}$  & 252 & 29  & \textcolor{gray}{47.5} \% & \textcolor{gray}{4}\%    \\
GAT & 0.27  & 0.10            & 82  & 0    & 50\%    & 61.0\%   \\
GT &  0.47  & $5\cdot10^{-9}$     & 18193 & 1053   & \textbf{99.3}\% & \textbf{85.5}\%  \\
\hline
\end{tabular}
\caption{Ansatz-reduction performance for the $15\!\to\! 16$ experiment.}
\label{tab:ansatz_reduction}
\end{table}
Threshold selection could be coupled to the linear solve: start from a candidate threshold, and decrease it if the reduced ansatz yields no solution. An analysis of the efficiency gain of such a pipeline is left for the future. Here we quantify pruning: with optimal threshold GAT and GT remove up to $61\%$ and $85.5\%$ of the redundant $d$-graph sector respectively (millions of graphs), with GT improving from $42.6\%$ at $n+1=13$ to $85.5\%$ at $n+1=16$.
\paragraph{Interpretability}
We extract graph embeddings and train linear probes to predict pre-computed graph-level features~\cite{alain2017understanding}, testing whether global structure is encoded in the representations. 
We consider standard cycle counts $C_3,C_4,C_5$ and induced 4-node graphlets $\gOneFourTikz,\gThreeFourTikz,\gFourFourTikz$. Note that the cycles $C_n$ can include chords, additional edges across the cycle, so for example \gThreeFourTikz ~indicates the induced 4-cycle graphlet only, in contrast to  $C_4$. 
Table~\ref{tab:linear_probe_new} reports the corresponding $r^2$ scores.

\begin{table}[h!]
\centering
\small
\setlength{\tabcolsep}{4pt}
\begin{minipage}[c]{0.06\linewidth}
\centering
\end{minipage}\hfill
\begin{minipage}[t]{0.34\linewidth}
\centering
GAT\par\vspace{2pt}
\begin{tabular*}{\linewidth}{@{\extracolsep{\fill}}cccc@{}}
\toprule
 & $13\!\to\!14$ & $14\!\to\!15$ & $15\!\to\!16$  \\
\midrule
$C_3$      & \textbf{0.99} & \textcolor{gray}{0.23} & \textbf{0.99} \\
$C_4$      & \textbf{0.99} & \textcolor{gray}{0.17} & \textbf{0.99} \\
\gOneFourTikz  & 0.97 & \textbf{0.97} & \textbf{0.99} \\
\gThreeFourTikz  & 0.97 & 0.93 & 0.97 \\
\gFourFourTikz  & \textbf{0.99} & \textbf{0.97} & \textbf{0.99} \\
$C_5$      & 0.97 & 0.96 & 0.98 \\
\bottomrule
\end{tabular*}
\end{minipage}\hfill
\begin{minipage}[t]{0.30\linewidth}
\centering
GIN\par\vspace{2pt}
\begin{tabular*}{\linewidth}{@{\extracolsep{\fill}}ccc@{}}
\toprule
$13\!\to\!14$ & $14\!\to\!15$ & $15\!\to\!16$ \\
\midrule
0.76 & \textcolor{gray}{0.13} & 0.59 \\
\textbf{0.82} & \textcolor{gray}{0.10} & 0.59 \\
0.69 & \textcolor{gray}{0.42} & \textcolor{gray}{0.41} \\
0.76 & \textbf{0.63} & \textcolor{gray}{0.39} \\
0.79 & \textcolor{gray}{0.45} & 0.54 \\
0.79 & \textbf{0.63} & \textbf{0.64} \\
\bottomrule
\end{tabular*}
\end{minipage}\hfill
\begin{minipage}[t]{0.30\linewidth}
\centering
GT\par\vspace{2pt}
\begin{tabular*}{\linewidth}{@{\extracolsep{\fill}}ccc@{}}
\toprule
$13\!\to\!14$ & $14\!\to\!15$ & $15\!\to\!16$ \\
\midrule
\textbf{0.96} & \textcolor{gray}{0.15} & \textbf{0.96} \\
\textbf{0.96} & \textcolor{gray}{0.11} & \textbf{0.96} \\
\textbf{0.96} & 0.89 & \textbf{0.96} \\
0.95 & 0.85 & 0.95 \\
\textbf{0.96} & \textbf{0.90} & \textbf{0.96} \\
0.95 & 0.86 & 0.95 \\
\bottomrule
\end{tabular*}
\end{minipage}
\caption{$r^2$ values for predicting global $k$-cycles and 4-node graphlet counts from model embeddings.}
\label{tab:linear_probe_new}
\end{table}

Three features stand out. First, GAT embeddings achieve the highest $r^2$, indicating strong aggregation of local structure. Second, the  $14\!\to\!15$  experiment shows a striking anomaly: the predictability of $C_3$ and $C_4$ collapses, while induced graphlets and $C_5$ remain highly recoverable. The origin of this effect is unclear and points to a non-trivial structural transition in the dataset. Finally, the high $r^2$ for $C_5$, despite no explicit input, indicates that the embeddings capture higher-order, non-local structure.

\section{Discussion / Future work}
\label{sec:discussion}

Our results show that GNNs can identify redundant sectors of the graphical bootstrap while preserving exactness. Working at the $d$-graph level limits the reduction, capturing only a subset of vanishing $f$-graphs (34\% at $n=16$). A natural next step is $f$-graph-level pruning, which remains a binary task, deciding whether $c^{(n)}_{a,b}=0$, i.e. on graphs with both numerator and denominator edges. Direct coefficient prediction is a separate and harder problem; nevertheless, at fixed loop order the target is a finite and highly structured set of rational numbers~\cite{Bourjaily:2025iad}. At 13-loops we expect them to be multiples of $\frac{1}{8}$ ranging from $-42$ to $14$ (stronger constraints can also be conjectured). The coefficient-prediction problem could be then formulated as a structured finite-label task. Finally, we believe this work provides a useful benchmark for learning on purely combinatorial graphs and strengthens the connection between exact results in high-energy physics and machine learning. Code and data are available at \href{https://github.com/GabrieleDian/gnns-for-bootstrap}{code} and \href{https://huggingface.co/datasets/Gabriele-dian/graphical-bootstrap-correlator-dataset}{dataset}.

\begin{ack} 
This research was supported in part through the Maxwell computational resources operated at Deutsches Elektronen-Synchrotron DESY, Hamburg, Germany. The authors would like to thank Roi Naveiro Flores for insightful conversation and the beginning of the project. GD is grateful to Nicola Segala for stimulating discussions on graph spectra. PH is
supported by STFC Consolidated Grant ST/X000591/1. RA and GD
are funded by the Deutsche Forschungsgemeinschaft (DFG, German Research
Foundation) under Germany’s Excellence Strategy – EXC 2121 ”Quantum Universe“ – 390833306.
\end{ack}

\bibliographystyle{JHEP}
\bibliography{bibliography}

@article{dersy2022simplifying,
  author = {Dersy, Arnaud and others},
  title = {Simplifying Polylogarithms with Machine Learning},
  journal = {arXiv preprint arXiv:2206.04115},
  year = {2022}
}

@article{Eden:2000mv,
    author = "Eden, B. and Schubert, C. and Sokatchev, E.",
    title = "{Three loop four point correlator in N=4 SYM}",
    eprint = "hep-th/0003096",
    archivePrefix = "arXiv",
    reportNumber = "LAPTH-786-2000",
    doi = "10.1016/S0370-2693(00)00515-3",
    journal = "Phys. Lett. B",
    volume = "482",
    pages = "309--314",
    year = "2000"
}

@article{alnuqaydan2023symba,
  author = {Alnuqaydan, Mohammed and Gleyzer, Sergei and Prosper, Harrison B.},
  title = {SYMBA: Symbolic Computation of Squared Amplitudes in High Energy Physics with Machine Learning},
  journal = {Machine Learning: Science and Technology},
  volume = {4},
  number = {1},
  pages = {015007},
  year = {2023},
  doi = {10.1088/2632-2153/acb160}
}

@article{cai2024transforming,
  author = {Cai, Xinyi and others},
  title = {Transforming the Bootstrap: Using Transformers to Compute Scattering Amplitudes in Planar N = 4 Super Yang-Mills Theory},
  journal = {Machine Learning: Science and Technology},
  volume = {5},
  number = {3},
  pages = {035073},
  year = {2024},
  doi = {10.1088/2632-2153/ad6f61}
}

@article{cleveland2023graphlet,
    author = "Cleveland, Colin and Lee, Chin-Yen and Tsai, Shen-Fu and Yu, Wei-Hsuan and Lee, Hsuan-Wei",
    title = "{Graphlet and Orbit Computation on Heterogeneous Graphs}",
    journal = "arXiv preprint arXiv:2304.14268",
    year = "2023"
}

@article{Eden:2011we,
    author = "Eden, Burkhard and Heslop, Paul and Korchemsky, Gregory P. and Sokatchev, Emery",
    title = "{Hidden symmetry of four-point correlation functions and amplitudes in N=4 SYM}",
    eprint = "1108.3557",
    archivePrefix = "arXiv",
    primaryClass = "hep-th",
    reportNumber = "CERN-PH-TH-2011-208, DCPT-11-33, IPHT-T11-91, LAPTH-030-11",
    doi = "10.1016/j.nuclphysb.2012.04.007",
    journal = "Nucl. Phys. B",
    volume = "862",
    pages = "193--231",
    year = "2012"
}

@article{Eden:2012tu,
    author = "Eden, Burkhard and Heslop, Paul and Korchemsky, Gregory P. and Sokatchev, Emery",
    title = "{Constructing the correlation function of four stress-tensor multiplets and the four-particle amplitude in N=4 SYM}",
    eprint = "1201.5329",
    archivePrefix = "arXiv",
    primaryClass = "hep-th",
    reportNumber = "CERN-PH-TH-2012-014, DCPT-12-03, HU-EP-12-03, HU-MATH-2012-27, LAPTH-005-12, IPHT-T12-005",
    doi = "10.1016/j.nuclphysb.2012.04.013",
    journal = "Nucl. Phys. B",
    volume = "862",
    pages = "450--503",
    year = "2012"
}

@article{Bourjaily:2015bpz,
    author = "Bourjaily, Jacob L. and Heslop, Paul and Tran, Vuong-Viet",
    title = "{Perturbation Theory at Eight Loops: Novel Structures and the Breakdown of Manifest Conformality in N=4 Supersymmetric Yang-Mills Theory}",
    eprint = "1512.07912",
    archivePrefix = "arXiv",
    primaryClass = "hep-th",
    reportNumber = "DCPT-15-75",
    doi = "10.1103/PhysRevLett.116.191602",
    journal = "Phys. Rev. Lett.",
    volume = "116",
    number = "19",
    pages = "191602",
    year = "2016"
}

@article{Bourjaily:2016evz,
    author = "Bourjaily, Jacob L. and Heslop, Paul and Tran, Vuong-Viet",
    title = "{Amplitudes and Correlators to Ten Loops Using Simple, Graphical Bootstraps}",
    eprint = "1609.00007",
    archivePrefix = "arXiv",
    primaryClass = "hep-th",
    reportNumber = "DCPT-16-31",
    doi = "10.1007/JHEP11(2016)125",
    journal = "JHEP",
    volume = "11",
    pages = "125",
    year = "2016"
}

@article{He:2024cej,
    author = "He, Song and Shi, Canxin and Tang, Yichao and Zhang, Yao-Qi",
    title = "{The cusp limit of correlators and a new graphical bootstrap for correlators/amplitudes to eleven loops}",
    eprint = "2410.09859",
    archivePrefix = "arXiv",
    primaryClass = "hep-th",
    doi = "10.1007/JHEP03(2025)192",
    journal = "JHEP",
    volume = "03",
    pages = "192",
    year = "2025"
}

@article{Bourjaily:2025iad,
    author = "Bourjaily, Jacob L. and He, Song and Shi, Canxin and Tang, Yichao",
    title = "{Four-point correlator of planar supersymmetric Yang-Mills theory at twelve loops}",
    eprint = "2503.15593",
    archivePrefix = "arXiv",
    primaryClass = "hep-th",
    doi = "10.1103/kl4q-mpwp",
    journal = "Phys. Rev. D",
    volume = "112",
    number = "12",
    pages = "126029",
    year = "2025"
}

@article{xu2019powerful,
  title={How Powerful are Graph Neural Networks?},
  author={Xu, Keyulu and Hu, Weihua and Leskovec, Jure and Jegelka, Stefanie},
  journal={International Conference on Learning Representations (ICLR)},
  year={2019}
}

@inproceedings{xu2018representation,
  title={Representation Learning on Graphs with Jumping Knowledge Networks},
  author={Xu, Keyulu and Li, Chengtao and Tian, Yonglong and Sonobe, Tetsuya and Kawarabayashi, Ken-ichi and Jegelka, Stefanie},
  booktitle={International Conference on Machine Learning (ICML)},
  pages={5453--5462},
  year={2018}
}

@article{velivckovic2018graph,
  title   = {Graph Attention Networks},
  author  = {Veli{\v{c}}kovi{\'c}, Petar and Cucurull, Guillem and Casanova, Arantxa and Romero, Adriana and Li{\`o}, Pietro and Bengio, Yoshua},
  journal = {International Conference on Learning Representations (ICLR)},
  year    = {2018}
}

@article{Ying2021DoTR,
  title={Do Transformers Really Perform Bad for Graph Representation?},
  author={Chengxuan Ying and Tianle Cai and Shengjie Luo and Shuxin Zheng and Guolin Ke and Di He and Yanming Shen and Tie-Yan Liu},
  journal={ArXiv},
  year={2021},
  volume={abs/2106.05234},
  url={https://api.semanticscholar.org/CorpusID:235376852}
}

@inproceedings{vaswani2017attention,
  title={Attention Is All You Need},
  author={Vaswani, Ashish and Shazeer, Noam and Parmar, Niki and Uszkoreit, Jakob and Jones, Llion and Gomez, Aidan N. and Kaiser, {\L}ukasz and Polosukhin, Illia},
  booktitle={Advances in Neural Information Processing Systems (NeurIPS)},
  year={2017}
}

@misc{duong2019nodefeaturesgraphneural,
      title={On Node Features for Graph Neural Networks}, 
      author={Chi Thang Duong and Thanh Dat Hoang and Ha The Hien Dang and Quoc Viet Hung Nguyen and Karl Aberer},
      year={2019},
      eprint={1911.08795},
      archivePrefix={arXiv},
      primaryClass={cs.LG},
      url={https://arxiv.org/abs/1911.08795}, 
}

@misc{dwivedi2022graphneuralnetworkslearnable,
      title={Graph Neural Networks with Learnable Structural and Positional Representations}, 
      author={Vijay Prakash Dwivedi and Anh Tuan Luu and Thomas Laurent and Yoshua Bengio and Xavier Bresson},
      year={2022},
      eprint={2110.07875},
      archivePrefix={arXiv},
      primaryClass={cs.LG},
      url={https://arxiv.org/abs/2110.07875}, 
}

@article{mcgraw1992common,
  title={A common language effect size statistic.},
  author={McGraw, Kenneth O and Wong, S Paul},
  journal={Psychological Bulletin},
  volume={111},
  number={2},
  pages={361--365},
  year={1992},
  publisher={American Psychological Association},
  doi={10.1037/0033-2909.111.2.361}
}

@inproceedings{alain2017understanding,
  title={Understanding intermediate layers using linear classifier probes},
  author={Alain, Guillaume and Bengio, Yoshua},
  booktitle={International Conference on Learning Representations (ICLR)},
  year={2017}
}

@article{Henn:2020omi,
    author = "Henn, Johannes M.",
    title = "{What Can We Learn About QCD and Collider Physics from N=4 Super Yang{\textendash}Mills?}",
    eprint = "2006.00361",
    archivePrefix = "arXiv",
    primaryClass = "hep-th",
    doi = "10.1146/annurev-nucl-102819-100428",
    journal = "Ann. Rev. Nucl. Part. Sci.",
    volume = "71",
    pages = "87--112",
    year = "2021"
}
\end{document}